\documentclass[prb,amsmath,amssymb]{revtex4}% Physical Review B

\usepackage{graphicx}% Include figure files
\usepackage{dcolumn}% Align table columns on decimal point
\usepackage{bm}% bold math
\usepackage{color}

\newcommand\Tm{\langle\mathbf{T}\rangle}

\begin{document}

\title{Probing microscopic origins of confined subdiffusion by first-passage observables.}

\author{S. Condamin}
\affiliation{Laboratoire de Physique Th\'eorique de la Mati\`ere Condens\'ee
(UMR 7600), case courrier 121, Universit\'e Paris 6, 4 Place Jussieu, 75255
Paris Cedex}
\author{V.Tejedor}
\affiliation{Laboratoire de Physique Th\'eorique de la Mati\`ere Condens\'ee
(UMR 7600), case courrier 121, Universit\'e Paris 6, 4 Place Jussieu, 75255
Paris Cedex}

\author{R. Voituriez}
\affiliation{Laboratoire de Physique Th\'eorique de la Mati\`ere Condens\'ee
(UMR 7600), case courrier 121, Universit\'e Paris 6, 4 Place Jussieu, 75255
Paris Cedex}

\author{O. B\'enichou$^\star$}
\affiliation{Laboratoire de Physique Th\'eorique de la Mati\`ere Condens\'ee
(UMR 7600), case courrier 121, Universit\'e Paris 6, 4 Place Jussieu, 75255
Paris Cedex}

\author{J. Klafter}
\affiliation{School of Chemistry, Tel Aviv University, Tel Aviv 69978, Israel}

\date{\today}

\maketitle
$^\star$ To whom correspondence should be addressed. E-mail: benichou@lptmc.jussieu.fr

\textbf{Classification}\newline
 PHYSICAL SCIENCES, physics
\vspace{1cm}

\textbf{Corresponding author}\newline
Olivier B\'enichou\newline
LPTMC, Universit\'e Paris 6\newline
case courrier 121, 4 Place Jussieu, 75255 Paris Cedex 05\newline
Tel: + 33 1 44 27 25 29\newline
Fax: + 33 1 44 27 51 00\newline
benichou@lptmc.jussieu.fr

\vspace{1cm}

\textbf{Manuscript information}\newline
14 text pages, 1 table, 3 figures.

\newpage

\textbf{Subdiffusive motion of tracer particles in complex crowded environments, such as biological cells, has been shown to be widepsread. This deviation from brownian motion is usually characterized by a sublinear time dependence of the mean square displacement (MSD). However,   subdiffusive behavior can stem from  different microscopic scenarios, which  can not be identified solely by the  MSD data. In this paper we present a theoretical framework which permits  to calculate analytically first-passage observables  (mean first-passage
times,  splitting probabilities and   occupation times distributions) in disordered media in any dimensions. This analysis is applied to two representative microscopic models
of subdiffusion:  continuous-time random walks with heavy tailed waiting times, and  diffusion on  fractals. Our results show that first-passage observables provide tools to  unambiguously  discriminate between  the two possible microscopic scenarios of subdiffusion. Moreover we suggest experiments based on first-passage observables  which could help in determining  the origin of
subdiffusion  in complex media such as living cells, and discuss the implications of anomalous transport to reaction kinetics in cells.}

\newpage

\textbf{\Large Introduction}

In the last few years, subdiffusion has been observed in an increasing number
of systems\cite{Metzler00,R.Metzler2004}, ranging from physics\cite{Scher1975,Kopelman1984} or  geophysics\cite{Scher2002}  to biology\cite{Tolic-Norrelykke2004,Golding2006}. In particular, living cells provide striking examples for systems where subdiffusion has been repeatedly observed experimentally, either  in the cytoplasm\cite{Tolic-Norrelykke2004,Golding2006,Caspi2000,Yamada2000},  the nucleus\cite{Wachsmuth2000,Platani2002} or  the plasmic membrane\cite{KUSUMI1993,GHOSH1994,Smith1999}. However, the microscopic origin of subdiffusion in cells remains debated, even if believed to be due to crowding effects in a wide sense as indicated by in vitro experiments\cite{Amblard1996,Legoff2002,Wong2004,Banks2005}.

The subdiffusive behavior significantly deviates from the usual Gaussian solution of the simple diffusion equation, and is usually characterized by a mean square displacement (MSD)
which scales as \cite{Metzler00}
$\langle \Delta{\bf r}^2\rangle \sim t^\beta$ with $\beta < 1$.
Such a scaling law can be obtained from a few  models based on
different underlying microscopic mechanisms. Here we focus on two possibilities\footnote{ A third classical model of subdiffusion is given by the Fractional Brownian Motion which concerns processes with long range correlations}: (i) the first class of models that we consider stems from
continuous time random-walks (CTRWs)\cite{Metzler00,J.Klafter1987} and their continuous limit described by fractional diffusion
equations\cite{Metzler00,W.R.Schneider1987}. The anomalous behavior in these models originates from a heavy tailed distribution of waiting
times\cite{Sokolov2006}: at each step the walker lands on a trap, where it can be trapped for extended periods of  time. When dealing with  a tracer particle,  traps can be out-of-equilibrium chemical binding configurations\cite{Saxton1996,Saxton2007}, and the waiting times  are then the dissociation times;  traps can  also be realized by the free cages around the tracer in a hard sphere like crowded environment, and   the waiting times are  the life times of the cages (see figure 1a).
(ii) Another kind of model for subdiffusion relies on spatial inhomogeneities as
exemplified by diffusion in deterministic or
random fractals such as critical  percolation clusters \cite{BenAvraham,Bunde1991,1983}. The anomalous behavior is in this case due to the presence of fixed obstacles\cite{SAXTON1994} which create  numerous dead ends, as illustrated by De Gennes's ``ant in a labyrinth''\cite{degennesant}  (see figure 1b). These two scenarios can be classified as dynamic (CTRW) and static (fractal) in the nature of the underlying environment.

While these two  models lead to similar scaling laws for the
MSDs, their microscopic origins are intrinsically different and lead to notable differences in   other transport properties. This has strong implications, in particular on transport limited reactions\cite{Lomholt2007}, which will prove to  have very different kinetics in the two situations. As most of  functions of a living cell are regulated by coordinated chemical reactions which involve  low concentrations of reactants (such as transcription factors or vesicles carrying targeted proteins\cite{alberts}), and which are  limited by transport, understanding the origin of anomalous transport in cells and its impact on reaction kinetics is an  important issue.

Here we describe and analytically calculate the following transport related observables, based on first-passage properties, which allow as shown below to  discriminate between the CTRW and fractal models, and permits a quantitative analysis of the kinetics of transport limited reactions:

(a) The  first-passage time (FPT), which is the time needed for a particle starting from site $S$ to reach a target $T$ for the first time.
This quantity is fundamental in the study of transport limited reactions
\cite{Rice1985,S.B.Yuste2002,natphys}, as it gives the reaction time in the limit of perfect reaction.
 This quantity is
also useful in target search problems \cite{SlutskyBiophys04,nousprotein,nousanimaux,Benichou2006,Eliazar2007,Kolesov2007}, and
other physical systems \cite{nature2007,nv2007,RednerBook}. We will be interested in both the probability density function (PDF) of the FPT, and its first moment, the mean FPT (MFPT).

(b) The  first-passage
splitting probability, which is the probability to reach a target $T_1$ before
reaching another target $T_2$, in the case where several targets
are available. This quantity permits to study quantitatively competitive reactions\cite{Rice1985}.

(c) The occupation time before reaction, which is the time spent by a particle at a given site $T_1$ before reaction with a target  $T_2$. This quantity is useful in the context of reactions occurring with a finite probability per unit of time\cite{Blanco03,B2005,Condamin2007b}. We stress that the occupation time provides a finer information on the trajectory of the particle. In particular the FPT is given by the sum over all sites of the occupation time.  We will be interested in both the entire PDF of the occupation time, and the mean occupation time.

On the theoretical level, our approach permits the direct evaluation of  non trivial first-passage characteristics of transport in disordered media in any dimensions, while so far mainly effective one-dimensional geometries have been investigated \cite{RednerBook}. In particular we calculate here for the first time the MFPT, splitting probabilities and occupation time distribution of a random walk on percolation clusters, and discuss the potential implications of these results on reactions kinetics in living cells. We further argue  that our findings could lead to an experimental probing of the microscopic origin of subdiffusion in complex media like cells.

The paper is organized as follows. In the first section, we set the theoretical framework and give explicit analytical expressions of the first-passage observables, which are  summarized in equations (\ref{mfpt},\ref{sp},\ref{scalingt}). We then apply these results to the two above mentioned models of subdiffusion, namely the diffusion on fractal and CTRW models. In the second section, we discuss the relevance of these two models to describe anomalous transport in complex media like living cells, and suggest experiments which could help discriminating the microsopic origin of subdiffusion.

$ $

\textbf{\Large Results}

\textbf{Theoretical framework.}  Using recent techniques developed in (\cite{Condamin2005a,Condamin2007,nature2007}), we derive general analytical expressions of the first-passage observables. We consider a Markovian random walker
moving in a bounded domain of size $N$ with reflecting walls.
Let $W({\bf r},t|{\bf r}')$ be the propagator, i.e. the  probability density to be
at site ${\bf r}$ at time $t$, starting from the site ${\bf r}'$ at time $0$, whose evolution is described by a master equation\cite{VanKampen}
\begin{equation}
\frac{\partial W}{\partial t}={\cal L}W
\end{equation}
with a given transition operator $ {\cal L}$. We denote by  $P({\bf r},t|{\bf r}')$ the probability density that the first-passage time
to reach  ${\bf r}$, starting from ${\bf r}'$, is $t$. For the sake of simplicity we assume that the walker performs symmetric jumps, so that the stationary distribution is homogeneous $\lim_{t\to\infty} W({\bf r},t|{\bf r}')=1/N$. The propagator and first-passage time densities  are known to be related through \cite{Hughes}
\begin{equation}
W({\bf r}_T,t|{\bf r}_S) = \int_0^t P({\bf r}_T,t'|{\bf r}_S)
W({\bf r}_T,t-t'|{\bf r}_T) dt'.
\end{equation}
Following (\cite{nature2007}), this equation gives an exact expression for the MFPT, provided it is
finite:
\begin{equation}\label{Tm}
\Tm = N(H({\bf r}_T|{\bf r}_T)-H({\bf r}_T|{\bf r}_S)),
\end{equation}
where $H$ is the pseudo-Green function\cite{Barton1989}  of the domain :
\begin{equation}\label{pseudo}
H({\bf r}|{\bf r}') = \int_{0}^{\infty} (W({\bf r},t|{\bf r}') - 1/N)
dt.
\end{equation}

It is also possible to compute splitting probabilities
within this framework. If the random walker can be absorbed either
by a target $T_1$ at ${\bf r}_{1}$, or a target $T_2$ at ${\bf r}_{2}$,
a similar calculation yields:
\begin{equation}
\Tm/N =
P_1 H({\bf r}_1|{\bf r}_1) - P_2 H({\bf r}_1|{\bf r}_2) +
H({\bf r}_1|{\bf r}_S),
\end{equation}
where $P_1 $ (resp. $P_2$) is the splitting probability
to hit $T_1$ (resp. $T_2$) before $T_2$ (resp. $T_1$), and
$\Tm$ is the mean time needed to hit any of the targets.
This equation together with
the similar equation obtained by inverting $1$ and $2$, and  the
condition $P_1 + P_2 = 1$, give a linear system of 3 equations for the 3 unknowns $P_1$, $P_2$, and $\Tm$, which can therefore be straightforwardly determined. In particular the splitting probability $P_1$ reads:

\begin{equation}
\displaystyle P_1 = \frac{H_{1s}+H_{22}-H_{2s}-H_{12}}{H_{11}+H_{22}-2H_{12}},
\label{splitting}
\end{equation}
where we used the  notation $H_{ij} = H({\bf r}_i|{\bf r}_j)$.
This formula extends a previous result \cite{Condamin2005a,Condamin2007} obtained for simple random walks to the case of general Markov processes.

Beyond their own interest, the splitting probabilities allow us to obtain  the entire
distribution of the occupation time\cite{Condamin2007b} $\mathbf{N}_i$ at site $i$ for general Markov processes.
Denoting  $P_{ij}(i|S)$ the  splitting probability to reach
$i$ before $j$,  starting from $S$,   we have $P({\bf N_i} =0)= P_{iT}(T|S)$, and  for $k \geq 1$:
\begin{equation} \label{1} P({\bf N_i}=k) = E_1E_2(1-E_2)^{k-1},
\end{equation} where
\begin{equation} \label{E1} E_1 \equiv P_{iT}(i|S) =
\frac{H_{iS}+H_{TT}-H_{ST}-H_{iT}}{H_{ii}+H_{TT}-2H_{iT}},
\end{equation} and $E_2$ is the probability to reach $T$ starting from $i$ without ever returning to $i$ which reads\cite{Condamin2007b}:
\begin{equation}\label{E2}
E_2  =  \frac{1}{H_{ii}+H_{TT}-2H_{iT}}.
\end{equation}
In particular, the mean occupation time is then given by
\begin{equation} \langle {\bf N_i} \rangle =H_{iS} - H_{iT} + H_{TT}
- H_{ST}.
\label{resni}
\end{equation}
We stress that equation (\ref{1}) gives the exact distribution of the occupation time for all regimes. It follows in particular that the large time asymptotics of the occupation time distribution is exponential. Actually one can argue in the general case that the FPT is also exponentially distributed at long times. This comes from the fact that the transition operator ${\cal L}$ has a strictly negative discrete spectrum for a finite volume $N$ (see (\cite{VanKampen})).

 Equations (\ref{Tm},\ref{splitting},\ref{resni}) give exact expressions of the first-passage observables as functions  of the pseudo-Green function $H$.  The key point is that as shown in (\cite{nature2007}),  $H$  can be satisfactorily approximated by its infinite space limit,  which is precisely the usual Green function $G_{0}$:
\begin{equation}\label{integral}
H({\bf r}|{\bf r}')\approx G_0({\bf r}|{\bf r}') = \int_0^\infty W_0({\bf r},t|{\bf r}') dt,
\end{equation}
where $W_0$ is the infinite space propagator.
Following (\cite{nature2007}), we assume that the problem is scale invariant and 
we use for $W_0 $ the standard scaling\cite{BenAvraham}  :
\begin{equation}
W_0({\bf r},t|{\bf r}') \sim t^{-d_f/d_w} \Pi\left(
\frac{|{\bf r}-{\bf r}'|}{t^{1/d_w}}\right),
\label{scaling}
\end{equation}
 where the fractal dimension $d_f$ characterizes the
accessible volume $V_r \sim r^d_f$ within a sphere of radius $r$,   and  the walk dimension $d_w$  characterizes  the distance $r \sim t^{1/d_w}$ covered by a random walker
in a given  time $t$. The form (\ref{scaling}) ensures the normalization of $W_0$ by integration over the whole fractal set. Note that the MSD is then given by $\langle \Delta{\bf r}^2\rangle \sim t^\beta$ with $\beta =2/d_w$.
A derivation  given in (\cite{nature2007}) then allows to extract the scaling of the pseudo-Green function $H$, and eventually yields for the MFPT:
\begin{equation}
\Tm \sim \left\{
\begin{array}{ll}
N(A - B r^{d_w-d_f}) & \; {\rm for}\; d_w<d_f\\
N(A + B \ln r) & \; {\rm for}\; d_w=d_f\\
BN r^{d_w-d_f} & \; {\rm for}\; d_w>d_f
\end{array}
\right.
\label{mfpt},
\end{equation}
where explicit expressions of $A$ and $B$ are given in (\cite{nature2007}). We stress that in  the case of compact exploration ($d_w>d_f$), the MFPT depends on a single constant $B$. Indeed, the constant $A$ introduced in (\cite{nature2007}) can be shown to be actually $0$ in this case of compact exploration in scale invariant media.
In fact, the above analysis of the pseudo-Green functions also permits  to obtain explicit expressions of the splitting probabilities and mean occupation times:
\begin{equation}
P_1 \sim \left\{
\begin{array}{ll}
\displaystyle \frac{A + B( r_{1S}^{d_w-d_f}-r_{2S}^{d_w-d_f}-r_{12}^{d_w-d_f})}{2(A - B r_{12}^{d_w-d_f})} & \; {\rm for}\; d_w<d_f\\
\displaystyle \frac{A + B\ln(r_{2S}r_{12}/r_{1S})}{2(A + B \ln(r_{12}))} & \; {\rm for}\; d_w=d_f\\
\displaystyle \frac{1}{2}\left( (r_{2S}/r_{12})^{d_w-d_f}-(r_{1S}/r_{12})^{d_w-d_f}+1\right) & \; {\rm for}\; d_w>d_f
\end{array}
\right.
\label{sp}
\end{equation}
and
\begin{equation}
\langle {\bf N_i} \rangle \sim \left\{
\begin{array}{ll}
\displaystyle A + B( r_{iS}^{d_w-d_f}-r_{iT}^{d_w-d_f}-r_{ST}^{d_w-d_f}) & \; {\rm for}\; d_w<d_f\\
\displaystyle A + B\ln(r_{iT}r_{ST}/r_{iS}) & \; {\rm for}\; d_w=d_f\\
\displaystyle B( r_{iT}^{d_w-d_f}+r_{ST}^{d_w-d_f}-r_{iS}^{d_w-d_f}) & \; {\rm for}\; d_w>d_f
\end{array}
\right.,
\label{scalingt}
\end{equation}
where $r_{ij}=|{\bf r}_i-{\bf r}_j|$ is  different from $0$. Note that the entire distribution of ${\bf N_i}$ is obtained similarly by estimating $E_1$ and $E_2$ as defined by equations (\ref{E1},\ref{E2}).
Strikingly, the constants $A$ and $B$ do not depend on the confining
domain and can be written solely in terms of the  infinite space scaling function $\Pi$. We point out that in the case of compact exploration the expression of the splitting probability is fully explicit and does not depend on  $\Pi$.
Equations (\ref{mfpt},\ref{sp},\ref{scalingt}) therefore elucidate the dependence of the first-passage observables on the geometric parameters of the problem, and constitute the central theoretical result of this paper. We discuss the implications of these results on explicit examples in the next paragraph.

$ $

\textbf{Diffusion on  fractal model.} Critical
percolation clusters (see figure 1b) constitute a representative example of
random fractals \cite{Havlin1987,Bunde1991,BenAvraham}.
Here we consider  the case of  bond percolation, where the bonds connecting
the sites of a regular lattice of the $d$--dimensional space are present with probability $p$. The ensemble
of points connected by bonds is called a cluster. If $p$ is above the
percolation threshold $p_c$, an infinite cluster exists. If $p = p_c$, this infinite
cluster is a  random fractal characterized by its fractal dimension $d_f$.   We  consider  a nearest neighbor random walk on such critical  percolation cluster, with the so--called ``blind ant\cite{Hughes}'' dynamics 
: on arrival at a given site $\bf{s}$, the walker attempts to move to one of the adjacent sites on the original lattice with equal probability. If the link corresponding to this move does not exist, the walker remains at site $\bf{s}$. This walk is characterized by the walk dimension $d_w$. In the example of the 3--dimensional cubic lattice, one has $d_f = 2.58...$, and $d_w = 3,88...$\cite{Bunde1991} and the motion is subdiffusive with $\beta=2/d_w\simeq0.51..$.  For a given critical percolation cluster, namely for a given configuration of the disorder, the theoretical development of previous paragraph holds, and the first-passage observables are given by the exact expressions (\ref{Tm},\ref{splitting},\ref{resni}). However,  the variations between different realizations of the
disorder have to be taken into account, and averaging  has to be  performed in order to obtain meaningful quantities. It is shown in the Materials and Methods section that expressions (\ref{Tm},\ref{splitting},\ref{resni}) actually still hold after disorder averaging.

 Figure (2a,b,c) shows that the simulations fit very well the expected scaling. Both the volume dependence and the source-target distance dependence are faithfully reproduced by our theoretical expressions, as shown by the data collapse of the numerical simulations.

If the bond concentration $p$ is above the percolation
threshold $p_c$,  a correlation length $\xi\propto(p-p_c)^{-\nu}$  appears, where $\nu=0.87..$ for $d=3$. At length scales smaller than $\xi$, the  percolation cluster is fractal, with
the same fractal dimension $d_f$ as  the critical percolation cluster, and diffusion is anomalous. At length scales larger than $\xi$, the fractal dimension of the  percolation cluster recovers the space dimension $d$ and diffusion is normal\cite{BenAvraham}.

Along the lines of the previous section, we thus expect the pseudo-Green function $H$ to scale
as $r^{d_w-d_f}$ for $r<\xi$, and as $r^{d-2}$ for $r>\xi$. More explicitly, on the example of the MFPT we expect for the 3--dimensional cubic lattice
\begin{equation}
\Tm \sim \left\{
\begin{array}{ll}
B N r^{1.36...} & \; {\rm for}\;r<\xi\\
N(A' - B'/r) & \; {\rm for}\; r>\xi
\end{array}
\right.
\label{mfptcross}.
\end{equation}
Similarly, the other first-passage observables display a cross-over between these two regimes around $\xi$.  The simulations do show very well the transition between the two
regimes (see figure (2d)).

$ $

\textbf{CTRW model.} The CTRW is not necessarily Markovian unlike the fractal case, and therefore the above methodology can not be applied directly. The distribution of the FPT for CTRWs has however been obtained recently in (\cite{Condamin2007a}). We here briefly recall these results, and derive analytical expressions  of the other observables.
The CTRW  is  a standard random walk with random waiting
times, drawn from a PDF $\psi(t)$.
The CTRW model has a normal diffusive behavior if the mean waiting time is
finite.  For heavy tailed distributions such that
\begin{equation}
\psi(t) \sim \frac{\alpha \tau^\alpha}{\Gamma(1-\alpha)t^{1+\alpha}}\;\; {\rm for}\; t\gg \tau,
\label{defpsi}
\end{equation}
the mean waiting time diverges for $\alpha<1$ and the walk  is subdiffusive
since the MSD scales like $\langle \Delta {\bf r}^2 \rangle \sim t^\beta$ with $\beta=\alpha$ (see (\cite{Metzler00,Scher1975})). Here $\tau$ is a characteristic  time in the process. We focus  on the  representative  case
of a  one-sided Levy stable distribution \cite{Hughes} $\psi(t)$, which
satisfies equation (\ref{defpsi}) and whose Laplace transform is $\hat{\psi}(u)= \exp(-\tau^\alpha u^\alpha)$ ($0<\alpha<1$).

We  now derive the relation between the FPT to the site
${\bf r}_T$, starting from ${\bf r}_S$
for the standard discrete-time random walk and the CTRW.
Denoting $\pi(t)$ the probability density of the FPT for the CTRW, and
$Q(n)$ the probability density of the FPT for the discrete-time random
walk, $n$ being the number of steps, one has
\begin{equation}
\pi(t) = \sum_{n=1}^\infty Q(n) \psi_n(t),
\label{pit}
\end{equation}
which is conveniently rewritten after Laplace transformation as
\begin{equation}\label{piu}
 \widehat{\pi}(u)=\widehat{Q}(e^{-n\tau^\alpha u^\alpha}),
\end{equation}
where $\widehat{Q}(z)=\sum_{n=1}^\infty Q(n)z^n$ is the generating function of the discrete-time random
walk.

Several comments are in order. (i) First, the small $u$ limit shows that the  the MFPT is infinite, and the long-time
behavior of $\pi(t)$ is directly related to the MFPT of the discrete-time simple
random walk:
\begin{equation}
\label{ctrwlaw}
\pi(t) \sim \frac{\alpha \tau^\alpha}{\Gamma(1-\alpha)t^{1+\alpha}}
\langle n \rangle.
\end{equation}
It should be noted that as soon as $\widehat{Q}(z)$ is exactly known (such as  for $d=3$ in the large $N$ limit, see (\cite{Condamin2007a})), the entire distribution of the FPT can be obtained.   (ii) Second, as splitting probabilities are time independent quantities, they are exactly identical for CTRW and standard discrete time random walks, and are therefore given by equation (\ref{sp}) with the space dimension $d$ and the walk dimension $d_w=2$. (iii) Third, the same decomposition as equations (\ref{pit},\ref{piu}) holds for the distribution $\pi_i(t_i)$ of the occupation time $t_i$ of site $i$, where the distribution of the ocupation time $F({\bf N}_i)$ for  the discrete-time random
walk has to be introduced. This yields
\begin{equation}
\label{ctrwot}
\pi_i(t) \sim \frac{\alpha \tau^\alpha}{\Gamma(1-\alpha)t^{1+\alpha}}
\langle {\bf N}_i \rangle.
\end{equation}
Interestingly, as $F({\bf N}_i)$ is explicitly given by equation (\ref{1}), the entire distribution of the occupation time can be derived.

We emphasize that a proper definition of the mean values of the first-passage observables (namely the MFPT and the mean occupation time) is provided by introducing a  truncated  distribution (with cut-off $t_c$) of waiting times in place of $\psi(t)$. As this allows to define a mean waiting time $\tau_m=C\int_0^{t_c}t \psi(t)dt$ (where $C$ normalizes the truncated PDF), the MFPT is then given by $\Tm=\tau_m\langle n \rangle$, and the mean occupation time reads $\langle t_i \rangle=\tau_m \langle {\bf N}_i \rangle$.

Note our results show that the first-passage observables scale with the geometric parameters $N$ and $r$ exactly as a simple random walk. Their scaling dependence is therefore given by equations (\ref{mfpt},\ref{sp},\ref{scalingt}), where $d_f$ is the space dimension $d$ and $d_w=2$.

$ $

\textbf{\Large Discussion}

We first discuss the relevance of the two models, CTRW and diffusion on fractals to describe anomalous transport in confined systems such as  the cytoplasm and membrane of living cells.
The cell is  known to be a  highly complex and inhomogeneous molecular assembly, composed of numerous constituents which may widely vary from one cell type to another. Here we wish to distinguish between  two types of effects on transport in  cellular medium. First, the overall density of free proteins and molecular aggregates is very high, be it in the cytoplasm or in the plasma membrane. In such crowded environment, a tracer particle is trapped in dynamic  ``cages'' whose life times are broadly distributed at high densities and leading to equation (\ref{defpsi}). This dynamic picture therefore fits the hypothesis of the CTRW model. Second, the cytoskeleton is made of semiflexible polymeric filaments (such as F--actin or microtubules), which can be branched and cross--linked by proteins. This scaffold therefore acts as fixed obstacles constraining the motion of the tracer. Moreover, the cytoplasm can be compartmentalized by lipid membranes which further constrain the tracer. Such environment with obstacles can be described in a first approximation by a static percolation cluster. How could one discriminate between these two mechanisms having markedly different physical origin?

The first-passage  observables derived earlier make it possible to distinguish between the two models of subdiffusion, as summarized in Table (1). (i) The
first-passage time  has a finite mean and exponential tail for the fractal model, while it has  an infinite mean and a power law tail in a CTRW model. Analyzing the tail of the distribution of the FPT therefore provides a first tool to distinguish the two models. As   experiments
can only find the first-passage  up to a certain time, we need to use the above mentioned truncated  means  to define the MFPT for CTRW. In this case the scaling of the MFPT for CTRW with the source target distance is the same as for a simple  random walk, and can be distinguished from the scaling of the MFPT on random fractals. These two scalings are strikingly different for $d=3$:  the CTRW performs a non compact exploration of space ($d_w=2<3=d$) leading to a finite limit of the MFPT at large source-target distance, while exploration is compact for a random walker on the percolation cluster ($d_w>d_f$) leading to a scaling $\propto r^{d_w-d_f}$ of the MFPT. We highlight that this feature could have very strong implications on reaction kinetics in cells. Indeed, in the cases where the fractal description of the cell environment is relevant, our results show that reaction times crucially depend on the source target distance $r$. The biological importance of such dependence on the starting point has been recently emphasized in (\cite{Kolesov2007}), on the example of gene colocalization. On the other hand, when the CTRW description of transport is valid, reaction times do not depend on the starting point at large distance $r$. (ii)  The splitting probabilities for the CTRW model and for the
fractal models have  different scalings with the distance
between the source and the targets. As mentioned previously the difference is more pronounced for $d=3$: the
probability to reach the furthest target $T_2$ vanishes as $r^{-(d_w-d_f)}$
for the fractal model, $r$ being the distance $ST_1$ with the notations of  figure 3, while it tends to a constant for $d=3$ according to the CTRW model. As discussed above, this could have important consequences for the kinetics of competitive reactions in cells. (iii) As for the occupation time, both its distribution and the scaling of the conditional mean with the distances $ST_1$ and  $ST_2$ can be used to distinguish between models. The advantage of the mean occupation time is that it can still discriminate between the models  after averaging over initial conditions, and could therefore be used even with a concentration of tracers.

We now briefly discuss  potential experimental utilizations of first-passage observables.
The schematic set-up that we propose to measure these observables relies on single particle tracking techniques (see figure 3). We consider a single tracer, either a fluorescent particle or a nanocrystal, moving in a finite volume such as a living cell,  a microfluidic chamber or vesicle. A laser excitation defines the starting zone $S$. As soon as the tracer enters $S$ a signal is detected and a clock is started. Similarly, a second laser excitation defines the target zone $T_1$, and allows the measurement of the FPT of the tracer at $T_1$. In the same way, a third laser can detect a second target $T_2$: counting the time spent by the tracer in $T_2$  before reaching $T_1$ gives exactly the occupation time. Splitting probabilities are straightforwardly deduced.

Finally, this theoretical framework can be extended to cover more realistic situations. First, subdiffusion could result in some systems from a combination of both the dynamic (CTRW) and static (diffusion on fractal) mechanisms. Interestingly, our approach can be adapted to study the example of CTRWs on a fractal which models such situations\cite{Blumen1984}. Indeed, the same decomposition as in equation (\ref{pit}) holds in this case and shows  that the dependence of the first-passage observables (defined with truncated means if needed) on the source-target distance is exactly the same as in the case of a standard discrete-time random walk on the fractal, and therefore gives access to the dimensions $d_w$ and $d_f$ of the fractal. In turn, the tail of the distribution of the FPT is in this case reminiscent of the single step waiting time distribution defining the CTRW as shown by equation (\ref{ctrwlaw}) (see also ref(\cite{Blumen1984})). First passage observables therefore permit in principle to isolate and characterize each of the CTRW and fractal mechanisms even when they are both involved simultaneously.  Second, in various systems subdiffusion occurs over a given time scale or
length scale,  crossing over to the regular diffusive behavior. Both models can be adapted to capture this effect. In the fractal model the  fractal structure persists up to the
crossover length scale (which is the correlation length $\xi$ in percolation clusters above criticality),
 and the waiting time distribution for the CTRW model has a Levy-like
decay until the crossover timescale, after which the decay is faster so that the
mean waiting time  becomes finite.
The MFPT will exist in both of these modified models, but the CTRW model leads to
a normal scaling of the MFPT with the volume and the source-target distance: namely , it corresponds to the results
of the simple random walk, with the same time step as the mean waiting time.
On the other hand, a truncated fractal  structure would lead to the
same scaling on larger scales, but to a scaling as in equation (\ref{scalingt})
at smaller scales. The small-distance behavior of the MFPT  can thus
discriminate the two models. The same conclusion holds for the splitting
probabilities and occupation times: the small-length behavior will also differ.

Our approach therefore permits to explore the scaling of first-passage observables for two representative models of subdiffusion as a methodology to discriminate between underlying mechanisms for subdiffusion and to gain insight into the microscopic origin of subdiffusion and the nature of transport limited reactions in complex systems.

We thank P. Desbiolles for useful discussions.

$ $

\textbf{\Large Materials and Methods}

\textbf{Disorder average in the diffusion on fractal model }. We will denote by $\bar X$ the average of $X$ over the disorder, and assume that all configurations have the same volume $N$, which is a non restrictive condition in the large $N$ limit since $N$ is self-averaging.  Equations (\ref{Tm},\ref{splitting},\ref{resni}) then show that averaging the first-passage observables amounts to averaging the pseudo-Green function, and therefore the propagator in virtue of (\ref{pseudo}). In the case of a random walk on a critical percolation cluster it has been shown  that the propagator
has a multifractal behavior \cite{Bunde1991}.
 This means that the propagator $W(r,t)$ has a very broad distribution, and is not self-averaging:
 its typical value is not its average value, which is dominated by rare events. In particular a scaling form of the averaged propagator is not available. However, this difficulty can be by--passed if one considers the  chemical distance $x$, i.e. the step length of the shortest path between two points. Indeed in the chemical space, the
 propagator does have a simple fractal scaling\cite{Bunde1991,Havlin1987} and  in the infinite volume limit the averaged propagator $\overline{W}_0(x,t)$ satisfies the scaling form (\ref{scaling}) (see (\cite{Bunde1991})). Note that this property is shared by most of random fractals \cite{Bunde1991}, and  makes the chemical distance space  a powerful tool to calculate disorder averages. The formalism derived in the previous section  can therefore be employed, and  the scaling
laws of the MFPT, splitting probability and mean occupation time \emph{averaged over the disorder}  are  given in chemical space by equations  (\ref{mfpt},\ref{sp},\ref{scalingt}), where $r$ is to be replaced by the chemical distance $x$. Note that in the chemical space, the fractal dimension is given by $d_f^c=d_f/d_{\rm min}$ and walk dimension is $d_w^c=d_w/d_{\rm min}$. The dimension $d_{\rm min}$ is the fractal dimension of chemical paths and permits to recover the  dependence on the euclidian distance $r$ through the scaling\cite{BenAvraham} $x\sim r^{d_{\rm min}}$, with $d_{\rm min} = 1.24..$ in the case of the three-dimensional cubic lattice\cite{BenAvraham}.

These scaling laws for the first passage observables can be tested numerically. We simulated in figure (2a,b,c) several critical percolation clusters on the three-dimensional cubic lattice
embedded in the confining domain, and we averaged for each set of chemical distances $\{x_{ij} \}$ the desired observable
 over all configurations of source and targets yielding the same set $\{x_{ij} \}$.

%\bibliographystyle{cj}
%\bibliography{../../biblio/biblioportable/liste}
%\bibliography{../../../biblio/liste}

\newpage

\newpage

\textbf{Legends of Table and Figures }

TABLE 1. Comparison of first-passage observables for CTRW and fractal models for $d=3$. For the cubic lattice $\beta\simeq1.3..$ and $C$ is a constant to be redefined on each panel.

FIG. 1. Two scenarios of subdiffusion for a tracer particle in crowded environments. {\bf a}: Random walk in a dynamic crowded environment. The tracer particle evolves in a cage whose typical life time diverges with density. This situation can be modeled by a CTRW with power-law distributed waiting times. {\bf b}: Random walk with static  obstacles.  This situation can be modeled by a random walk on a percolation cluster.

FIG. 2. Numerical simulation of first-passage observables for random walks on 3--dimensional  percolation
clusters. All the
embedding domains have reflecting boundary conditions. {\bf a}: MFPT for random walks on 3--dimensional critical percolation
clusters. For each size of the confining domain, the MFPT, normalized by the
number of sites $N$, is averaged both over the different target and starting
points separated by the corresponding chemical distance, and over percolation
clusters.  The black plain curve
corresponds to the prediction of equation (\ref{scalingt}) with $d^c_w-d^c_f\simeq  1$. {\bf b}: Splitting probability for random walks on 3--dimensional critical percolation clusters. The splitting probability $P_1$ to reach the target $T_1$ before the target $T_2$ is
averaged both over the different target  points $T_2$  and over the  percolation
clusters. The chemical distance $ST_1=10$ is fixed while the chemical distance $ST_2=T_1T_2$ is varied. The  black plain curve corresponds to the explicit theoretical expression (\ref{sp})
 with $d^c_w-d^c_f\simeq  1$. {\bf c}: Occupation time for random walks on critical percolation
clusters. For each size of confining domain, the occupation time of site $T_1$ before the target $T_2$ is reached for the first time is averaged  over the different target
points $T_2$  and over the percolation
clusters.  The chemical distance $ST_1=10$ is fixed while the chemical  distance $ST_2=T_1T_2$ is varied.
 The black plain curve
corresponds to the prediction of equation (\ref{scalingt}) with $d^c_w-d^c_f\simeq  1$. {\bf d}: the MFPT for  random walks on percolation clusters above
criticality for a $25\times25\times25$  confining domain. The MFPT, normalized by the number of sites $N$, is
averaged both over the different target and starting points separated by the
corresponding chemical distance, and over the percolation
clusters.

FIG. 3. Schematic proposed set-up to measure first-passage observables.

\newpage

\begin{center}
 \begin{tabular}{|c|c|c|}
\hline
& CTRW model & Fractal model \\
\hline

FPT distribution & $\propto 1/t^{\alpha+1} $ & $\propto e^{-Ct}$ \\
\hline
(Conditional) mean FPT & $\sim N(1-C/r)$

  &  $\sim CNr^{\beta}$\\
\hline
Splitting probability $P_1$ & $\displaystyle \sim \frac{1 + C( r_{1S}^{-1}-r_{2S}^{-1}-r_{12}^{-1})}{2(1 - C r_{12}^{-1})}$ & $\displaystyle \sim\frac{1}{2}\left( (r_{2S}/r_{12})^{\beta}-(r_{1S}/r_{12})^{\beta}+1\right) $\\
\hline
(Conditional) mean occupation time
 & $\displaystyle \sim  1 + C( r_{1S}^{-1}-r_{1T}^{-1}-r_{ST}^{-1})$ & $\displaystyle \sim C( r_{1T}^{\beta}+r_{ST}^{\beta}-r_{1S}^{\beta}) $  \\
$\langle {\bf N}_1 \rangle $  of site $T_1$ & & \\
\hline

 \end{tabular}

\end{center}
\vspace{1cm}
\begin{center}
Table 1.

\end{center}

\newpage

\begin{figure}
\centering\includegraphics[width = 0.6\linewidth,clip]{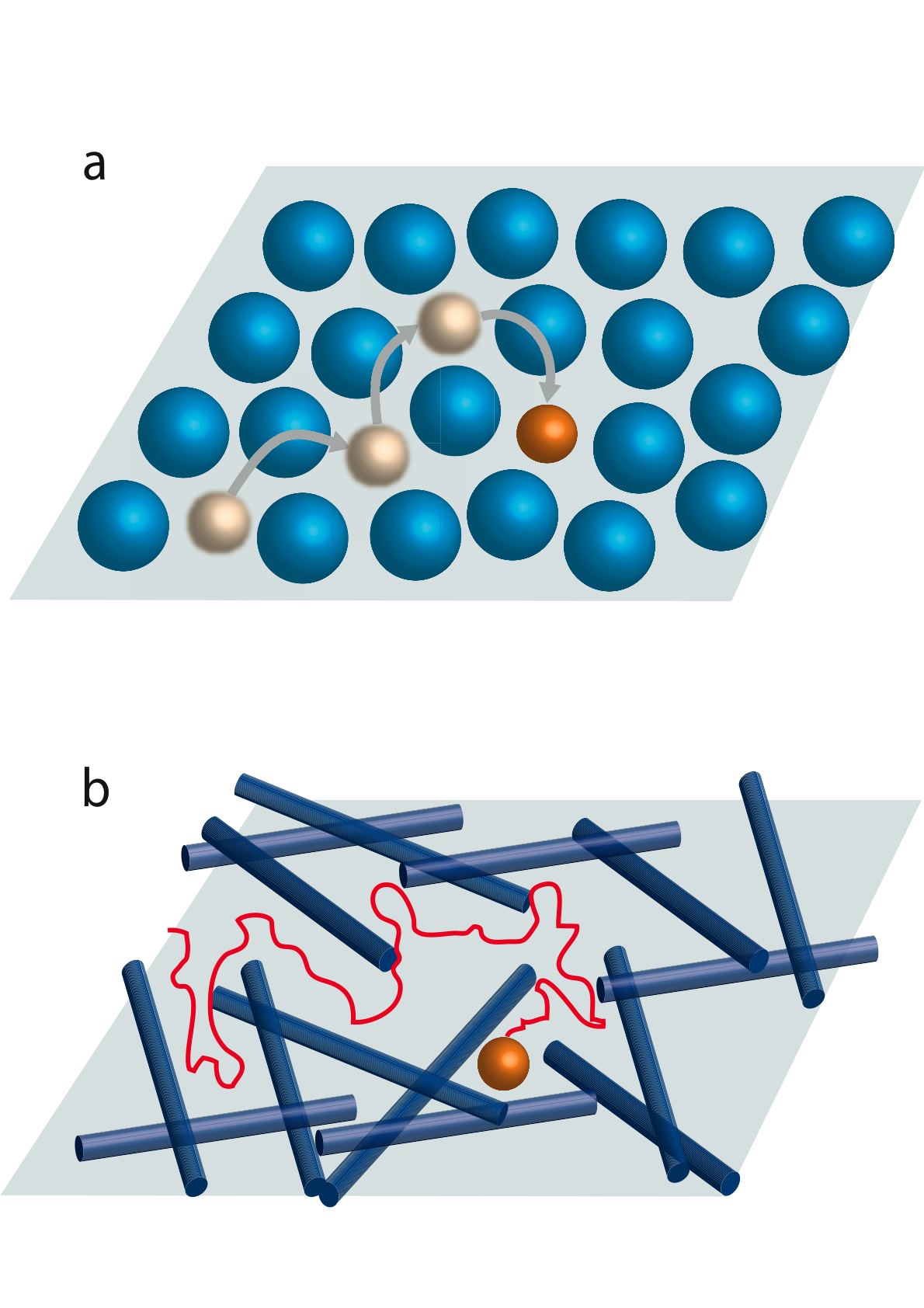}
\caption{}
\label{ctrw}
\end{figure}

\begin{figure}
\centering\includegraphics[width = .9\linewidth,clip]{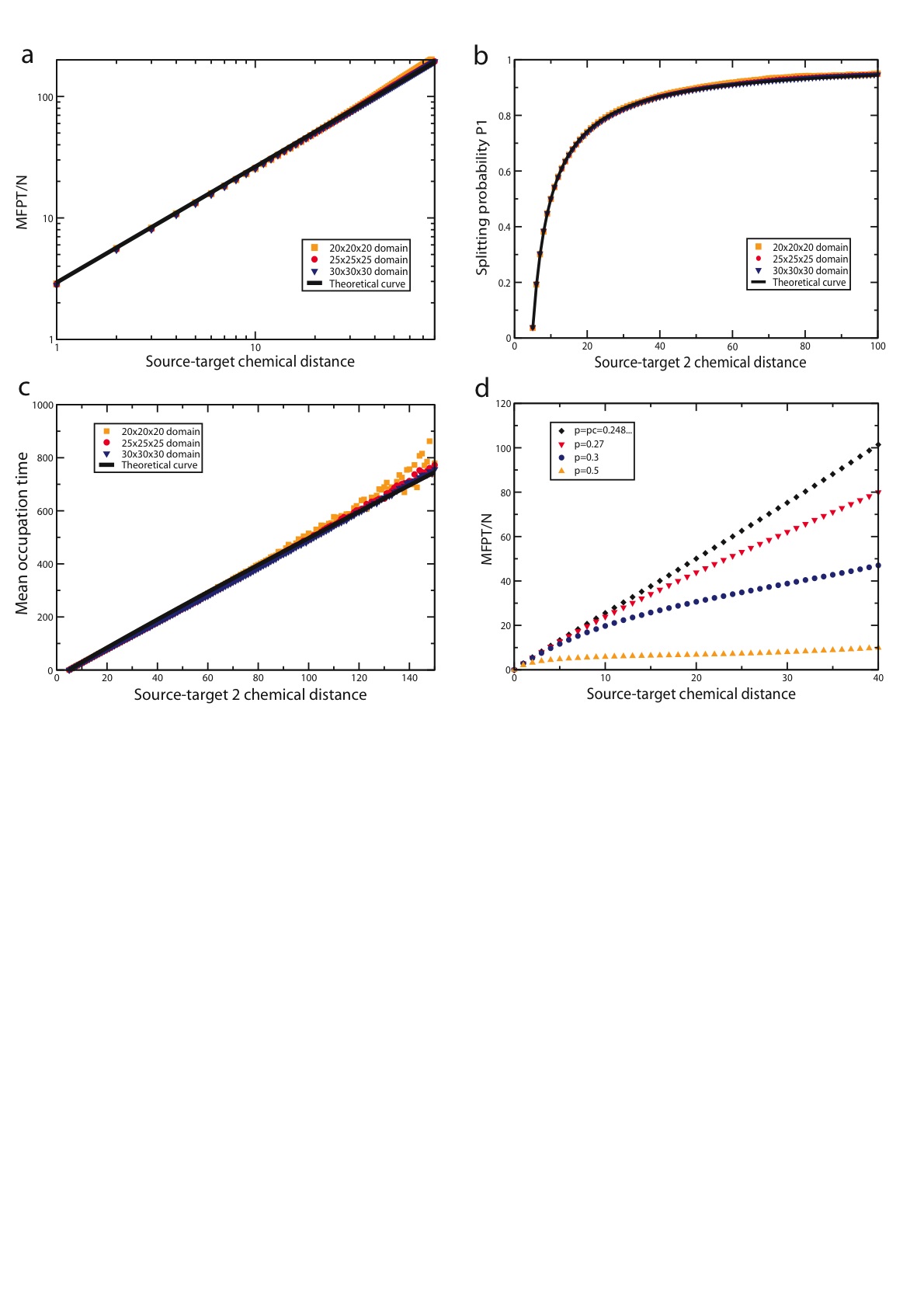}
\caption{}
\label{percolcrit}
\end{figure}

\begin{figure}
\centering\includegraphics[width = 1\linewidth,clip]{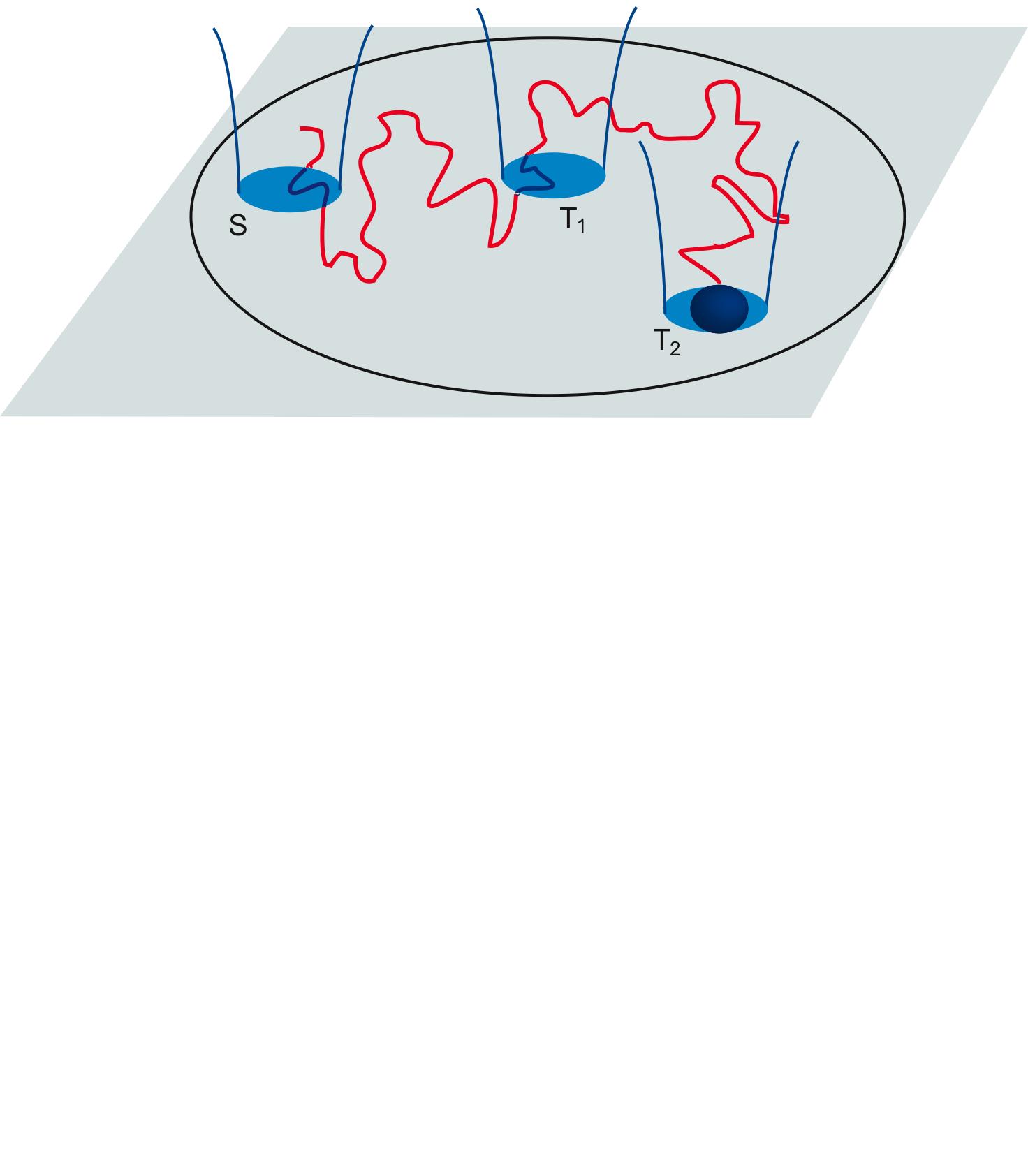}
\caption{}
\label{setup}
\end{figure}

\end{document}